      \documentclass[epj]{svjour}

\usepackage{graphicx}
\usepackage{epsfig}
\usepackage{amsmath}
\usepackage{amssymb}
\usepackage{amsfonts}
\usepackage{bm}

\begin{document}

\title{Short-time dynamics in the 1D long-range Potts model}
%  \subtitle{Do you have a subtitle?\\ If so, write it here}
\author{Katarina Uzelac\inst{1} \and
        Zvonko Glumac\inst{2}       \and 
        Osor S. Bari\v si\'c\inst{1}
       }% etc
%  \thanks{\emph{Present address:} Insert the address here if needed}%
% }                     % Do not remove
%Results
 \offprints{Katarina Uzelac}          % Insert a name or remove this line
  \mail{katarina@vrabac.ifs.hr}

\institute{Institute of Physics, POBox 304, Bijeni\v{c}ka cesta 46, HR-10001 Zagreb, Croatia
      \and Department of Physics, University J. J. Strossmayer, HR-31000 Osijek, Croatia }
\date{Received: date / Revised version: date}
\abstract{
   We present numerical investigations of the short-time dynamics at criticality in the 1D Potts model with power-law decaying interactions of the form  $1/r^{1+\sigma}$. The scaling properties of the magnetization, autocorrelation function and time correlations of the magnetization are studied. 
The dynamical critical exponents $\theta'$ and $z$ are derived in the cases $q=2$ and $q=3$ for several values of the parameter $\sigma$ belonging to the nontrivial critical regime.
\PACS{
      {05.50.+q}{Lattice theory and statistics}   \and
      {05.70.Jk}{Critical point phenomena}   \and
      {64.60.Ht}{Dynamic critical phenomena}   \and
      {61.20.Lc}{Time-dependent properties; relaxation} 
     } % end of PACS codes
} %end of abstract

\title{Short-time dynamics in the 1D long-range Potts model}

%\author{Katarina Uzelac$^1$, Zvonko Glumac$^2$, Osor S. Bari\v si\'c^1$}
%\email{katarina@ifs.hr}
%\email{zvonko.glumac@fizika.unios.hr}
%\email{obarisic@ifs.hr}

\maketitle

\section{Introduction}

	Short-time dynamics (STD) in systems quenched to criticality has attracted considerable attention in the last decade due to the appealing fact that systems even in the early period of relaxation to equilibrium exhibit universal scaling properties which involve both static and dynamic critical exponents \cite{JSS89,Huse}.
The interest in this phenomenon exists at different levels.
From a practical point of view, it offers a useful numerical tool for calculating both dynamic and  static critical properties where the critical slowing down is turned into advantage. 
From a fundamental point of view, it opened a series of questions of current interest 
 from the universal amplitudes to 
the universality of the fluctuation-dissipation ratio \cite{Cugliandolo93} in a wider context of ageing phenomena in pure systems \cite{CG05}.
One of the first points of conceptual interest was the emergence of a new independent universal dynamical exponent  describing the initial increase of the magnetization in this regime \cite{JSS89},
but related also to
the persistence probability of the global order parameter \cite{Majumdar96}.
Since the STD was formulated in the context of the dynamical renormalization group (RG) and the new exponent evaluated within the $\epsilon$-expansion \cite{JSS89} it has been further investigated, mostly numerically, in a variety of models in two and three dimensions for equilibrium phase transitions  \cite{SZ95,OSchYZ97,daSilva02,Zheng98} and also  for out-of-equilibrium ones \cite{TO98}.

Quite a few studies were carried out on models with long-range (LR) interactions.
The RG approach of Janssen et al. \cite{JSS89}  was extended to the case of power-law decaying interactions of the form $r^{-d-\sigma}$ in the same continuous n-vector model \cite{CGLMS00}, in the random Ising model \cite{Chen02},
and in the kinetic spherical model \cite{CGLY00,BDH07}.
Studies of STD at criticality in discrete models with LR interactions, where such an approach does not apply,  are still absent. 
Numerical "advantage" is there rather reduced due to the fast relaxation in the presence of LR interactions. 

In this paper we present the first and preliminary numerical study of the 1D LR Potts model, useful as a pa\-ra\-digm  that comprises different universality classes obtained by variation 
 of the number of states $q$.
We show that, in spite of the difficulties of the numerical approach in the LR case, the scaling properties characteristic for the STD may be well reproduced with a reasonable numerical effort and derive the two dynamical critical exponents in the wide extent of the range-parameter $\sigma$ for two different universality classes.  

The outline of the paper is as follows. In Section ~\ref{sec:Model} we give an overview of the model and basic STD properties considered in the paper, followed by the details of our numerical approach. 
The Section ~\ref{sec:Results} contains the results for two special cases of the Potts model: $q=2$, corresponding to the Ising model, which is compared to the previous RG results, and $q=3$, where the new results are derived in the regime where the transition is of the second order. 
The conclusion is given in Section ~\ref{sec:Conclusion}.

\section{Model and short-time dynamics approach}
\label{sec:Model}
	We consider the 1D Potts model defined by the Hamiltonian 
\begin{equation}
H = - \sum_{i < j} \; \frac{J}{|i - j|^{1+\sigma}} \; \delta _{s_i, s_j} \; ,
\label{hamilt}
\end{equation}
where $J>0$, $s_i$ denotes a $q$-state Potts spin at the site $i$, 
$\delta$ is the Kronecker symbol and the summation is over all the pairs of the system.
Hereafter $J=k_B=1$ is used.
As is well known \cite{ACCN88}, for $0 <\sigma \leq 1$ the model (\ref{hamilt}) has a phase transition at nonzero temperature for all $q$. 
Only a few exact results are available for its equilibrium critical behavior, but the model was studied in detail by several approximate methods \cite{GU93,LB97,BDD99}. 
It has a rather complicated phase diagram in the $(q,\sigma)$ plane, involving similar variety of critical regimes, that is  encountered in the $(q,D)$ plane of the same model with short-range (SR) interactions. This gives the additional motivation to examine also the dynamical scaling properties in the STD regime depending on $q$ and $\sigma$.

In the present work we are interested in two special cases, $q=2$ and $q=3$ in the range of parameter $\sigma$ corresponding to the nontrivial (non mean-field (MF)) critical regime, where the initial slip of the magnetization can be observed. For $q=2$ this is accomplished for $0.5 <\sigma <1$ \cite{FMN72}. In the latter case, $q=3$, which belongs to a different universality class, this region is restrained to $\sigma_{c}(q=3) < \sigma < 1$, where $\sigma_{c}(q) > 0.5$ denotes the point of the onset of the first-order phase transition, occurring for $q>2$ and known only approximately \cite{UG97GU98,ReynalDiep04}. 
For these two cases  we shall study the nonequilibrium evolution to criticality in early times of several quantities, magnetization, autocorrelation function and time correlations of the magnetization. 
Let us first briefly remind their scaling properties in the STD regime and explain their implementation to the model (\ref{hamilt}).

\subsection{STD approach}

As shown by Janssen {\it et al} \cite{JSS89}, if the system is brought out of equilibrium by a quench from high temperature to criticality, and left to evolve following the nonconservative dynamics of Model A (in the sense of reference \cite{HH77}), then, during the early stage of relaxation it will display universal scaling properties characterized by the static exponents and the new universal dynamic exponent. 
Consequently, in the system of size $L$ after a quench from high temperature to the critical region in the presence of small initial magnetization $m_0$, the magnetization will obey the scaling relation
\begin{equation}
M(t, \tau, L, m_0) = b^{-\beta/\nu} M(t/b^z, b^{1/\nu} \tau, L/b, b^{x_0}  m_0),
\label{scgen}
\end{equation}
where $\tau = (T-T_c)/T_c$, $b$ is a scaling factor and $\beta, \nu$ are the static critical exponents.
Besides the dynamical exponent $z$, the scaling involves a new exponent $x_0$ as the anomalous dimension of the initial magnetization $m_0$.

At criticality ($\tau = 0$), and for $L\gg \xi$,
%(after choosing $b=t^{1/z}$)
equation (\ref{scgen}) may be reduced to
\begin{equation}
M(t, m_0) = t^{-\beta/(\nu z)} M(1, t^{x_0/z}  m_0).
\label{scTc}
\end{equation}
For early times satisfying $t \ll t_x  \approx m_0^{-z/x_0}$, but larger than the microscopic time $t_{micro}$, the r. h. s. can be expanded giving the power-law increase of the magnetization known as the initial slip,
\begin{equation}
M(t)  \sim m_0   t^{\theta'},
\label{deftheta1}
\end{equation}
with $\theta' = x_0/z - \beta/(\nu z)$.
	The magnetization in the model (\ref{hamilt}) is defined in a standard way 
 \begin{equation}  \label{eq:m1}
M(t) =  \langle M_1(t) \rangle\; = \frac{q}{(q-1)\;L}\;\left<  \sum_{i}\;  \left( \delta_{s_i(t),1} - \frac{1}{q}   \right)  \right>, 
\end{equation}
where  $1$ denotes the preferential direction among $q$ possible Potts states ${\alpha}$.
The brackets $\langle...\rangle$ denote the average over initial conditions and random force. 

During the short time after the quench, the correlation length is small compared to the system size,
and the exponent $\theta'$ can be derived directly from the power law (\ref{deftheta1}) by performing simulations on the chain of a single large size and averaging over a great number of independent runs. 

In the absence of the initial magnetization ($m_0 = 0$), equation (\ref{scgen}) gives the scaling relation for the $k$-th moment of the magnetization, 
\begin{equation}
M^{(k)}(t,L) = b^{-k\beta/\nu} M^{(k)}(t/b^{z},L/b).
\label{eq:scMk}
\end{equation}
 In early times, when $\xi(t) \ll L$, the second moment also displays a power-law behavior, 
\begin{equation}
M^{(2)}(t,L) \sim t^{(d-2\beta/\nu)/z},
\label{eq:Mk}
\end{equation}
which can be used to derive the anomalous dimension of the order parameter $\beta/\nu$, or the dynamical exponent $z$ directly from the single large chain.
To this purpose we use the alternative definition of the order parameter
%(which remains nonzero even for finite systems)
\begin{equation}\label{eq:mx}
  M_x(t) =   \frac{q}{(q-1)\;L}\;  max_{\alpha} \left[ \sum_{i}\;  \left( \delta_{s_i(t),\alpha} - \frac{1}{q}   \right) \right]
%   M_x(t) =  \frac{max_\alpha[\frac{1}{L}\sum_{i}\;  (q \; \delta_{s_i(t),\alpha} - 1)]}{q - 1} 
\end{equation}
and the moments of magnetization are obtained as the average
\begin{equation}
 M^{(k)}(t) = \langle M^{k}_x(t) \rangle. 
\label{mxk}
\end{equation}
Equation (\ref{eq:mx}) describes the absolute value of the magnetization and allows us to apply the scaling relation  (\ref{eq:scMk}) already to the first momentum, and
obtain
\begin{equation}
\langle M_{x}(t) \rangle\; \sim \; t^{(d/2-\beta/\nu)/z}.
\label{eq:mxpl}
\end{equation}

The autocorrelation function of the local order parameter is defined in a standard way and also obeys the power-law form 
 \begin{equation}
A(t) = \frac{q}{(q-1)\;L}\; \left< \sum_{i}\;  \left(\delta_{s_i(0),s_i(t)} - \frac{1}{q}\right) \right> \;  \sim \;  t^{-\lambda/z},   
\label{autocorr}
\end{equation}
depending on the combination of both dynamical exponents  $\lambda/z=d/z - \theta'$.

For the calculation of the exponent $\theta'$ we shall use another quantity which represents 
the autocorrelation of the global order parameter.
It was shown by Tom\'e and de Oliveira \cite{TO98}  that the time correlation  
of magnetization defined as
\begin{equation}
Q(t) = \langle M_1(0)M_1(t) \rangle
\label{eq:defQ}
\end{equation}
 also exhibits the initial increase of the power-law form
\begin{equation}
 Q(t) \sim t^{\theta'}
\label{gcorr}
\end{equation}
even in absence of the imposed initial magnetization.
(Notice that in Equation (\ref{eq:defQ})  the definition (\ref{eq:m1}) of magnetization should be used and not its  absolute value.)
For numerical calculations of the exponent $\theta'$, Equation (\ref{gcorr}) has a technical advantage compared to the expression (\ref{deftheta1}), where the runs should be performed first for several values of the initial magnetization $m_0$ and than the extrapolation to the limit $m_0 \rightarrow 0$ taken in order to obtain the exponent $\theta'$.
In return, however, the fluctuations are more pronounced for $Q(t)$ and its calculation requires better statistics. 

\bigskip

\subsection{Numerical calculations}

	Monte Carlo simulations were done on finite chains with periodic boundary conditions by using
simple Metropolis dynamics. 
The system was quenched from a random configuration (high-temperature state) to criticality. 

  Unlike the earlier studies for the short-range Potts model in 2D, where the critical temperatures are known exactly, in the LR case only the approximate results are available. 
Satisfactory results for series of different values of $\sigma$ were obtained by the finite-range scaling (FRS) approach \cite{UG88}, cluster mean-field approach \cite{Monroe99}, or Monte Carlo calculations \cite{ReynalDiep04}. 
In the present study we use the values for $T_c$ obtained by the FRS \cite{GU93}.

Two approaches were examined - a direct derivation of exponents and a derivation from the finite-size scaling (FSS).
In the former approach where the exponents are calculated using a single large system, the correlation length has to be small compared to the system size during times which are taken into account in the evaluation of the power laws. 
Due to LR interactions, the correlation length increases much faster than in systems with short-range interactions.  
For illustration we supply here a rough estimate of the increase of the correlation length $\xi(t)$ calculated from the 
second moment of the spin-spin correlation function \cite{Brezin82} at the instant t, 
\begin{equation}
\xi^{2}(t) = \frac{ \sum_{l=1}^{L/2} \; l^2 \;  C(l,t)}{\sum_l \; C(l,t)},
\label{eq:defksi}
\end{equation}
where the correlation function $C(l,t)$ is given by
\begin{equation}
C(l,t) =  \frac{q}{(q-1)\;L}\; \left< \sum_i \; \left( \;\delta_{s_i(t)s_{i+l}(t)} - \frac{1}{q}\right)  \right>.
\label{defcorr}
\end{equation}
Summation in equation (\ref{eq:defksi}) runs only up to $L/2$ because of the periodic boundary conditions.
\begin{table}[htb]
 \caption{Number of steps elapsed before the correlation length (\ref{eq:defksi}) has reached half of its maximum value for $q=2$.}\label{tb:Table1}
\begin{center} 
\begin{tabular}{crrrrr} 
\hline\noalign{\smallskip}
$\sigma \backslash L$ ~ &100&400 & $1\,000$ & $3\,000$  \\
\noalign{\smallskip}\hline\noalign{\smallskip}
0.9 & 2  & 13  & 40   & 168 \\
0.8 & 1  & 7   & 20   & 54 \\
0.7 & 1  & 4  & 10   & 30 \\
0.6 & 0  & 3  & 6   & 15  \\
\noalign{\smallskip}\hline
\end{tabular} 
\end{center} 
 \end{table}
As shown in Table \ref{tb:Table1}, the correlation length increases very rapidly indeed, especially for lower values of $\sigma$. 
Values for $q=3$ are similar.
Consequently, in order to reach sufficiently long time intervals in the power-law regime, all the direct calculations were performed with chains of $3\,000$ sites. 
All the quantities  were averaged over $200\,000$ to $350\;000$  independent runs. 
Larger numbers of independent runs were used for smaller values of $\sigma$, where the fluctuations
are more pronounced. 
Finally, in the FSS approach small sizes ranging from $L=100$ to $L=400$ were compared.

\section{Results}
\label{sec:Results}
	 Systematic calculations in cases $q=2$ and $q=3$ were performed for four characteristic values of parameter of range $\sigma =~0.6,~0.7,~0.8,~0.9$. 

Increasing of $\sigma$ by moving away from the MF regime up to the limits of relevance of long-range interactions has similar effect in this 1D model as leaving the MF regime by lowering dimensionality down to the lower critical dimensionality in its SR analogue, and we expect to observe similar features. One of them is dependence of dynamical exponents $\theta'$ and $z$ on $\sigma$. 

The above choice of $\sigma$ allows to cover evenly the nontrivial critical regime $0.5 < \sigma < 1$ for $q=2$. 
In the case $q=3$ it covers both first- and second-order transition regimes, but the detailed analysis is 
focused on the region where the second-order phase transition is expected.

\subsection{Case $q=2$}

\subsubsection{Time correlations of the magnetization and the exponent $\theta'$}

The dynamical exponent $\theta'$, which in the SR analogue increases with decreasing of dimensionality \cite{JSS89}, in the present LR case should increase with $\sigma$, which is also in agreement with the RG results \cite{CGLMS00}.

\begin{figure}[h!!!]
\begin{center}
\includegraphics[scale = 0.45]{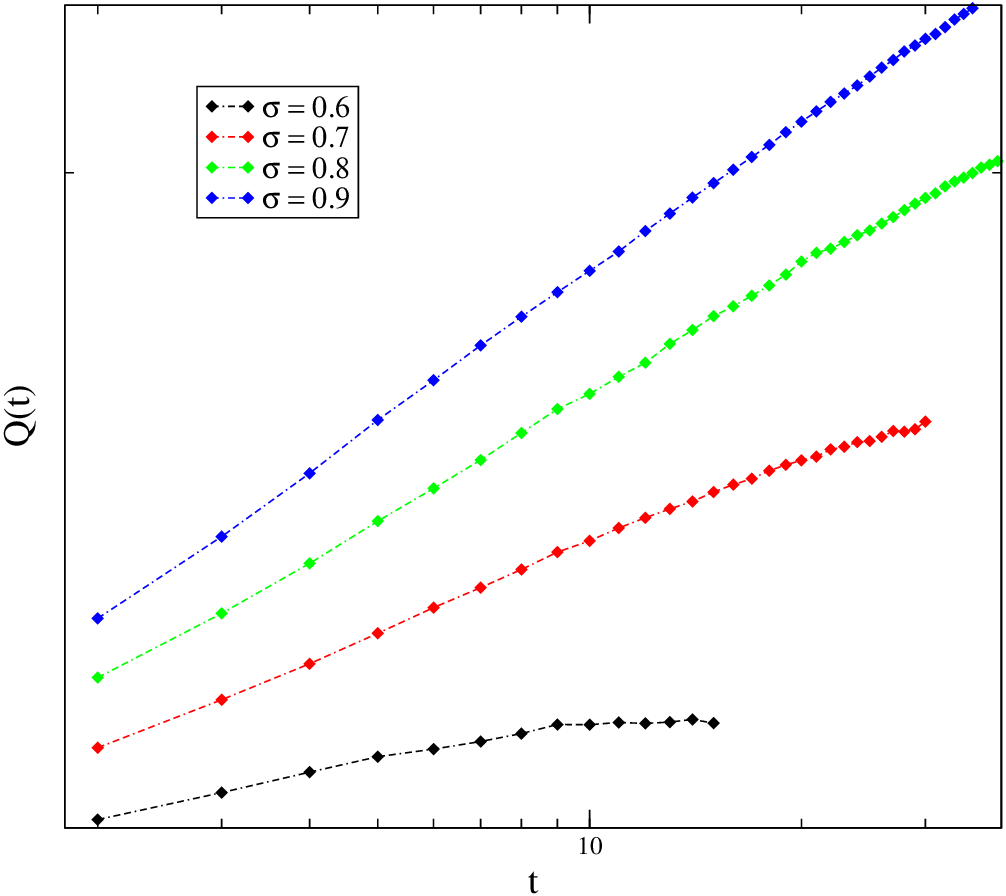}
\caption{The time correlations of the magnetization for $q=2$ and the selected values of $\sigma$ ($L=3\,000$).}
\label{fg:fig1}
\end{center}
\end{figure}

The principal quantity that we used to derive the exponent $\theta'$ is the function $Q(t)$ (\ref{eq:defQ}). 
A summary graph of our numerical simulations for the selected values of $\sigma$ is presented in Figure \ref{fg:fig1}.
%
% ------------ f i g u r e  1  -----------
% ---------------------------------------
One observes that the microscopic time $t_{micro}$ is short and the linear behavior on the logarithmic scale is established immediately after the first 2-3 steps.
The linear regime in the log-log scale becomes shorter as $\sigma$ decreases, but the size $L=3\,000$ is sufficiently long for an accurate evaluation of the exponent $\theta'$ by a fit to equation (\ref{gcorr}), which deteriorates only for the lowest $\sigma$ considered.

The errors in present results are not easy to estimate, because they may be introduced by several sources: insufficient statistics, arbitrariness in the selection of the linear segment of the plot, or using the approximate values for $T_c$.
The error bars given in tables cover the first two sources and could be systematically reduced by increasing the number of independent runs and the size of the chains considered. 
Yet, we stress that the third one cannot be estimated directly.

The obtained values for $\theta'$ are presented in Table \ref{tb:Table2} compared to the RG results by the two-loop $\epsilon$  expansion of reference \cite{CGLMS00}. 
 \begin{table}[htb]
   \caption{Dynamical exponents for $q=2$ compared to the RG results of reference \cite{CGLMS00} rounded up to 4 digits.}
 \label{tb:Table2}
 \begin{center} 
\begin{tabular*}{8.2cm}{@{}c@{}c@{}c@{}c@{}c@{}c@{}}
\hline\noalign{\smallskip}
$~\sigma$ & $\theta'$  & ~$\theta'_{RG}$  & ~$z$ & ~$z_{RG}$ & ~$\lambda/z$\\
\noalign{\smallskip}\hline\noalign{\smallskip}
0.9  &~ $0.212\pm .005$  &~ 0.3346  &~ $1.18\pm.04$   &~ 0.9532   &~ $0.635\pm .004$ \\
0.8  &~ $0.188\pm .004$  &~ 0.2587  &~ $0.96\pm.04$   &~ 0.8340   &~ $0.85~\pm .01~$ \\
0.7  &~ $0.137\pm .006$  &~ 0.1733  &~ $0.81\pm.01$   &~ 0.7174   &~ $1.136\pm .02~$ \\
0.6  &~ $0.07~\pm .01~$  &~ 0.0821  &~ $0.70\pm.01$   &~ 0.6052   &~ $1.47~\pm .02~$ \\
\noalign{\smallskip}\hline
\end{tabular*} 
  \end{center} 
 \end{table}
Since the accuracy of our results improves with increasing $\sigma$, one may conclude that the RG results are overestimated due to the insufficiency of the two-loop expansion in that regime. Similar overestimation of $\theta'$ was observed in the SR case, where the same RG $\epsilon$-expansion in the SR limit \cite{CGLMS00} gives e.g. $\theta'= 0.131$ and $\theta'= 0.356$  for $d=3$ and $d=2$ respectively, while the MC simulations give respectively $\theta'=0.104$ \cite{Grassberger} and $\theta'=0.191$ \cite{Grassberger,OSchYZ97}.

 A more standard way to calculate the exponent $\theta'$ is from the initial slip of the magnetization given by equation (\ref{deftheta1}). 
In the present problem we find it less advantageous both for precision and for the numerical effort needed.
For this reason we do not proceed with the systematic analysis using this approach. 
Just for illustration, we present the $\sigma = 0.9$ data in Table \ref{tb:Table3}, limiting ourselves to a very rough estimation.
\begin{table}
\caption{The exponent $\theta'$ calculated from the magnetization for $q=2$, $\sigma=0.9$ and $L=3\,000$ for several values of initial magnetization
         $m_0$ with the linear extrapolation to $m_0=0$.}
\label{tb:Table3}
\begin{tabular*}{8.2cm}{@{}c@{}c@{}c@{}c@{}c@{}}
\hline\noalign{\smallskip}
$~m_0~~$   &~~0.1 &~~0.05 & ~~0.01 &~~ $m_0 \rightarrow 0$  \\
\noalign{\smallskip}\hline\noalign{\smallskip}
$\theta'(m_0)$ &~~$.187\pm .005$ &~~ $.196\pm .006$ &~~$.202\pm .008$ &~~$.204\pm .009$ \\
\noalign{\smallskip}\hline
\end{tabular*}
\end{table}

The result is consistent with the one cited in Table \ref{tb:Table2}. Improving the accuracy would imply performing the calculations on several smaller initial values $m_0$, each of them requiring the same amount of numerical effort spent for the calculation of $Q(t)$.

\subsubsection{Magnetization and exponent $z$}

	As discussed earlier in Section 2.2, the dynamical exponent $z$ is expected to increase with $\sigma$, since the relaxation becomes slower with decreasing range of interactions. 
%Similar occurs with lowering dimensionality in the SR case.

\begin{figure}[hbt]
\begin{center}
\scalebox{0.45}{\includegraphics{{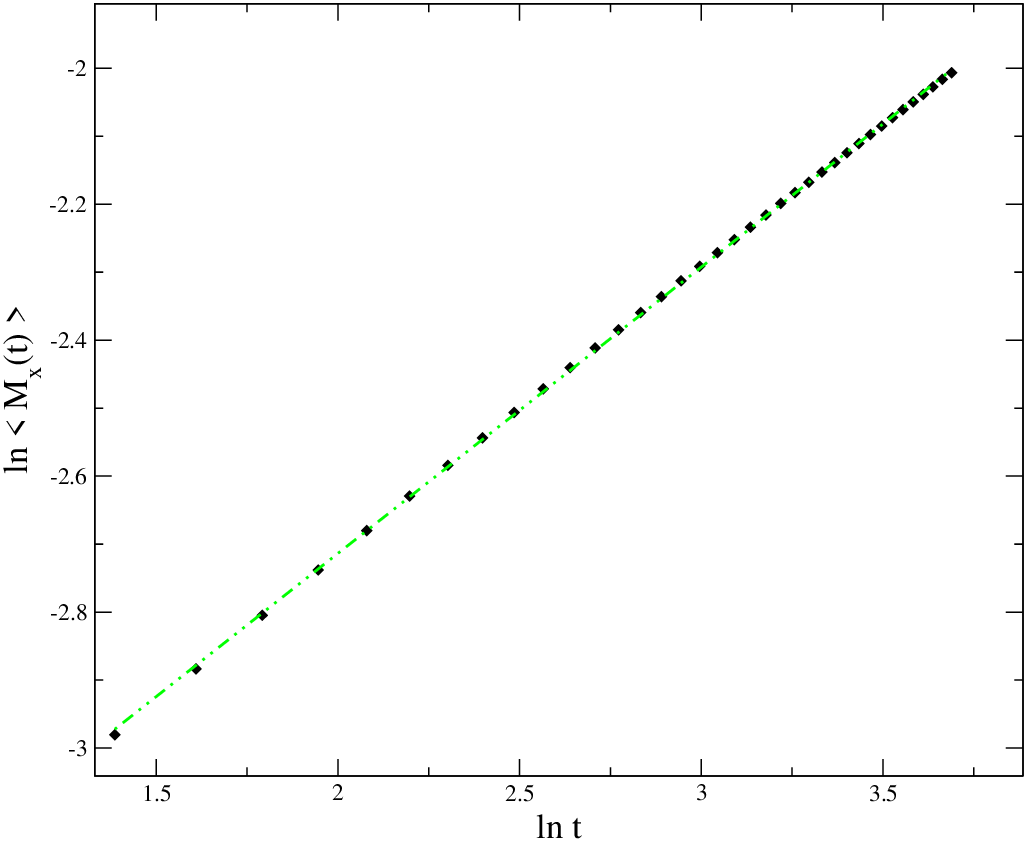}}}
\end{center}
\caption{Magnetization data for $q=2$, $\sigma=0.8$ (diamonds) and the linear extrapolation (dash-dotted line).} \label{fg:fig2}
\end{figure}

 	The exponent $z$ was calculated from the magnetization using equation (\ref{eq:mxpl}). 
The log-log plot of the simulation data  is illustrated in Figure \ref{fg:fig2} for 
%
% ------------ f i g u r e   -----------
% ---------------------------------------
the case $\sigma=0.8$.
The values of $z$ presented in Table \ref{tb:Table2} were obtained by substituting into equation (\ref{eq:mxpl}) the exact value for the anomalous dimension of the order parameter $\beta/\nu$, which is equal to $(1-\sigma)/2$  \cite{Brezin76}. 
As $\sigma$ increases our results become significantly larger than those obtained by the $\epsilon$ expansion \cite{CGLMS00}.
Again, we may attribute this discrepancy to an underestimation of the RG results by the two loop expansion and may observe similar behavior in the SR case, where, for the Ising model, the $(4-d)$-expansion to the second order \cite{HHM72} gives $z=2.013$ and $z=2.052$ for $d=3$ and $d=2$ respectively, while the best MC calculations give z close to 2.04 \cite{CG05} for d=3, and z=2.1667 \cite{NB00} for $d=2$.

\subsubsection{Autocorrelation function}

\begin{figure}[hbt!!!]
\begin{center}
{\scalebox{0.45}{\includegraphics{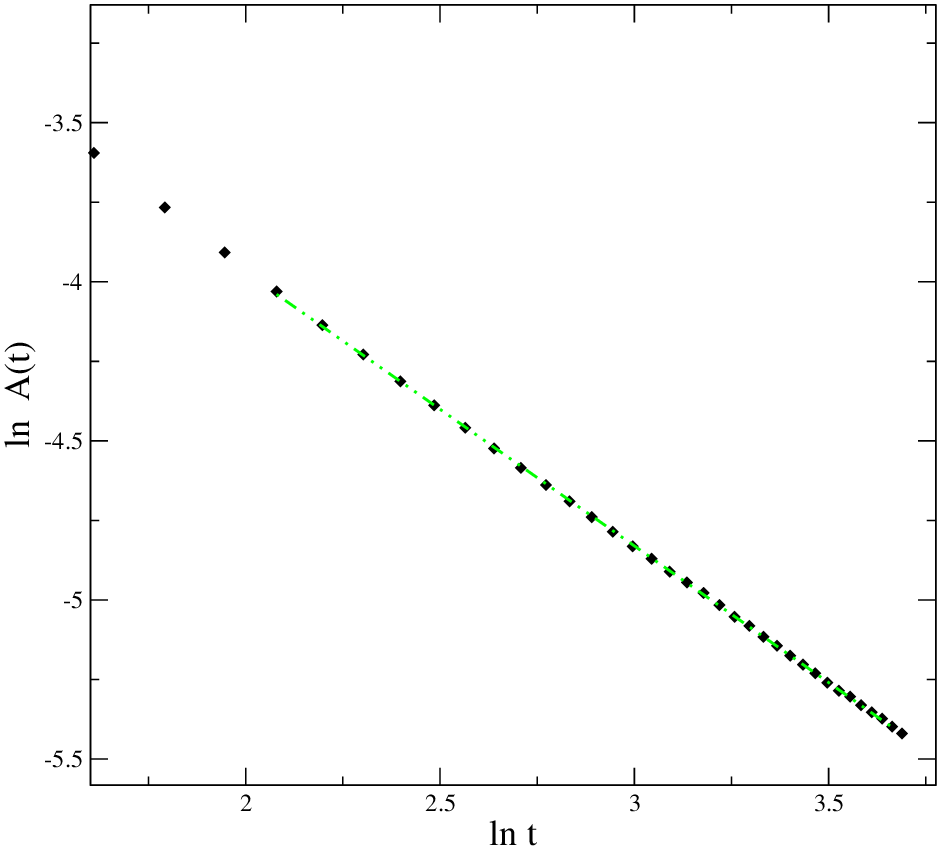}}}
\end{center}
\caption{MC data for autocorrelation function for $q=2$, $\sigma=0.8$, $L=3\,000$ (circles) and the linear extrapolation (dotted line).} 
\label{fig3}
\end{figure}

The example of simulations of the autocorrelation function for $\sigma = 0.8$ is illustrated by the log-log plots in Figure \ref{fig3}.
%
% ------------ f i g u r e   -----------
% ---------------------------------------
The power-law fit to equation (\ref{autocorr}) gives the exponent $\lambda/z$ presented in Table \ref{tb:Table2}.
Although not suitable for the calculation of the exponent $\theta'$ when the exponent $z$ is not known with sufficient precision, the values for $\lambda/z$ were used for a check of independent calculations of $\theta'$ and $z$.
Within given error bars, the agreement is obtained.

\subsubsection{Finite-size scaling}

An alternative way of evaluating the exponent $z$ is to perform the simulations on several small systems of different sizes and apply FSS by using the overlapping fits \cite{LSchZ96}. 
To this purpose one may consider the magnetization, Binder's fourth-order cumulant, but also the correlation length defined by equation (\ref{eq:defksi}).

\begin{figure*}
  \centerline{\includegraphics[height = 5cm, width = 7cm]{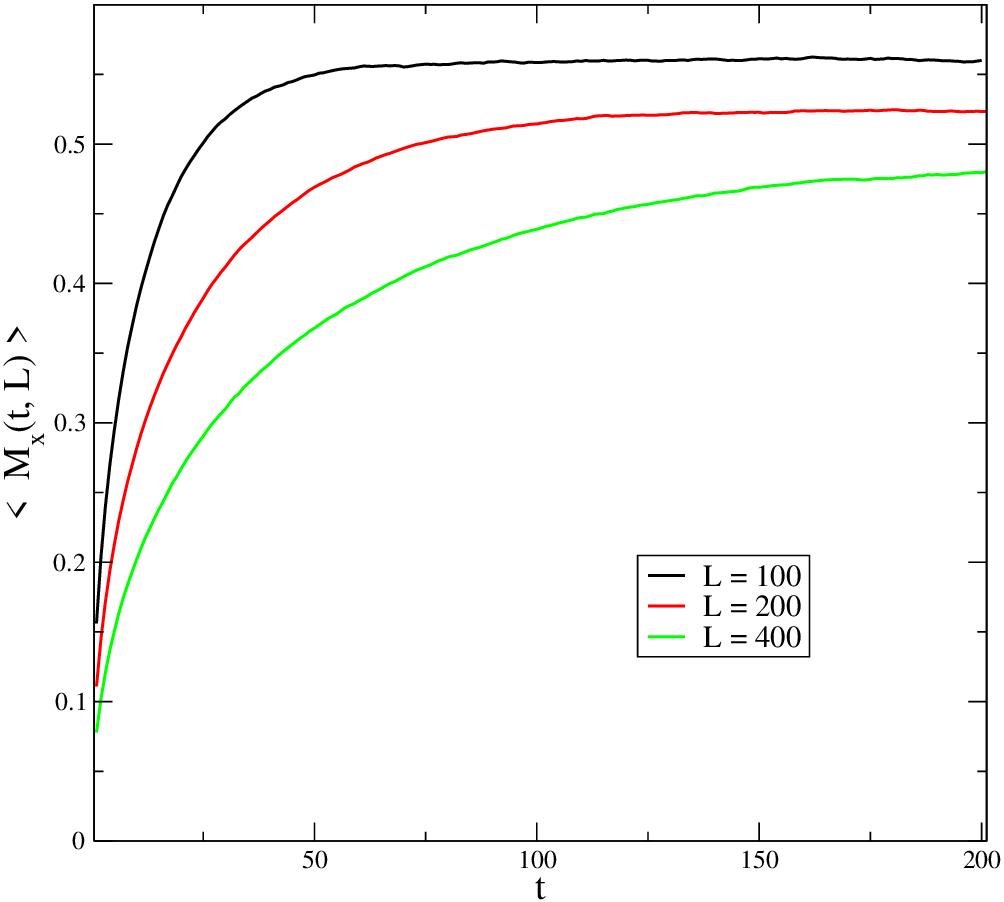}
              \includegraphics[height = 5cm, width = 7cm]{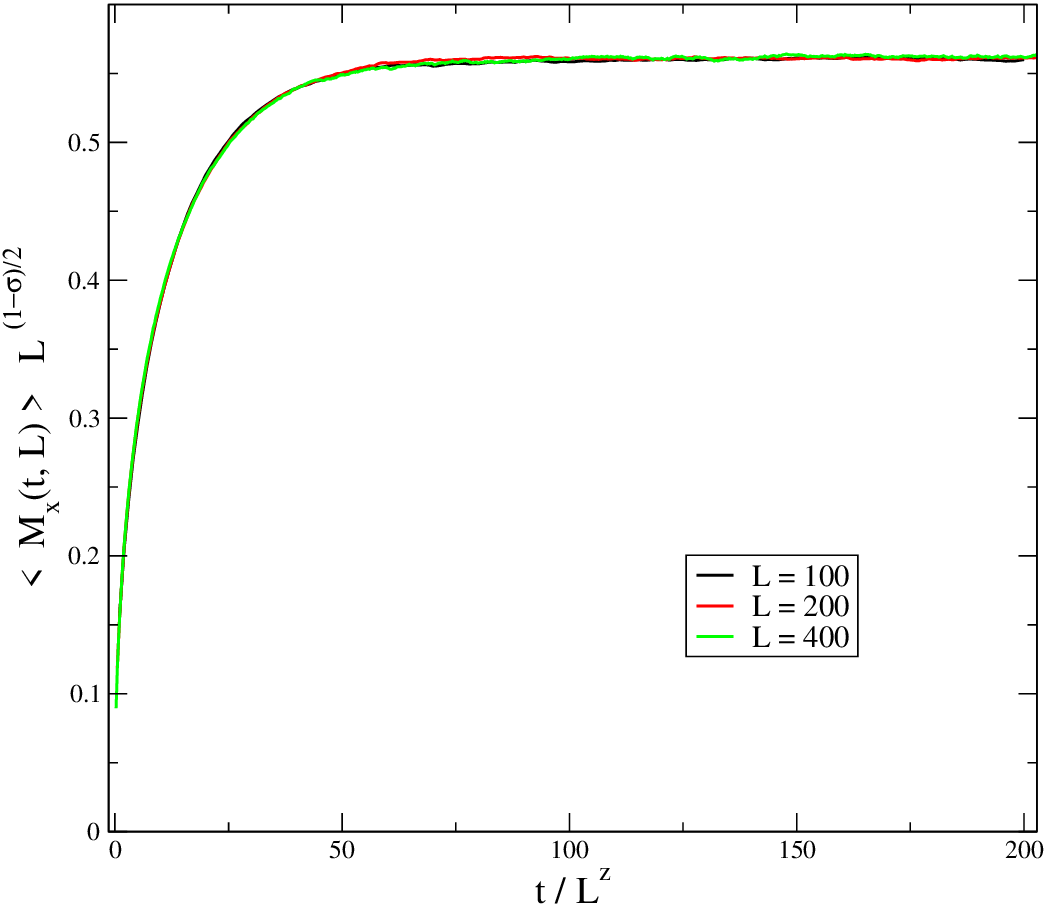}  }
% \vspace*{5cm}       % Give the correct figure height in cm
\caption{Scaled magnetization $M_x(t,L)$ for $L=100,200,400$ ($q=2$, $\sigma=0.8$). }
\label{fg:fig4}       % Give a unique label
\end{figure*}

\begin{figure*}
  \centerline{\includegraphics[height = 5cm, width = 7cm]{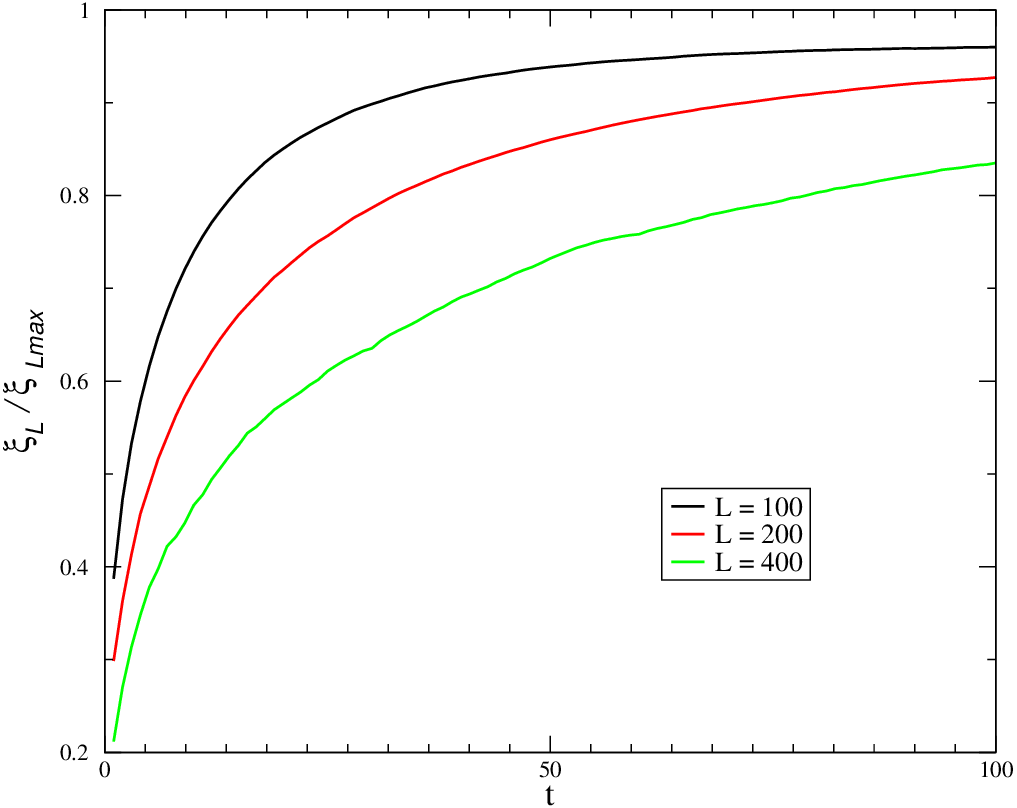}
              \includegraphics[height = 5cm, width = 7cm]{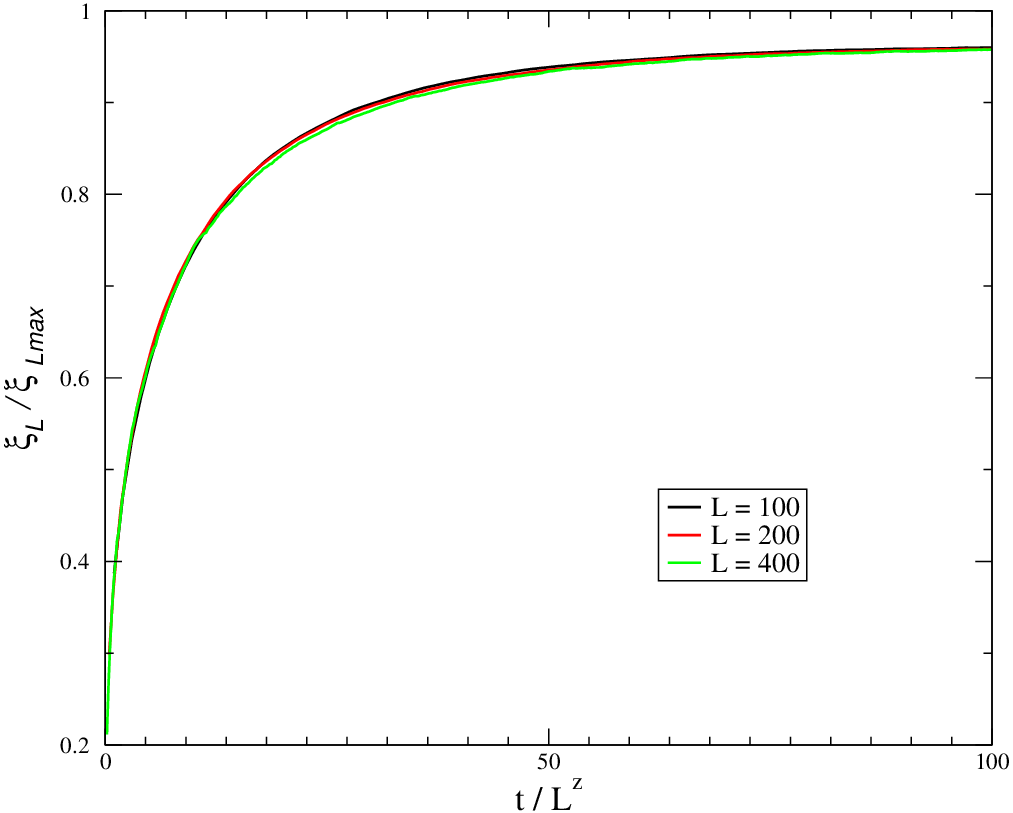}  }
% \vspace*{5cm}       % Give the correct figure height in cm
\caption{Scaled correlation length for $L=100,200,400$ ($q=2$, $\sigma=0.9$). }
\label{fg:fig5}       % Give a unique label
\end{figure*}

We illustrate two such fits, involving sizes $L=100, 200$ and 400, for the magnetization and the correlation length in Figures \ref{fg:fig4} and \ref{fg:fig5}.
% ------------ f i g u r e   -----------
%    \begin{figure}[htb]
%    \begin{center}
%    {\scalebox{0.4}{\includegraphics{fig4a.eps}}}
%    \centering\hspace{0.1cm}
%    {\scalebox{0.4}{\includegraphics{fig4b.eps}}}
%    \end{center}
%    \caption{Scaled magnetization $M_x(t,L)$ for $L=100,200,400$ ($q=2$, $\sigma=0.8$)}
%    \label{fg:fig4}
%    \end{figure}
% ---------------------------------------

The fit for the magnetization was performed by applying the scaling relation (\ref{eq:scMk}) to the magnetization defined by (\ref{eq:mx}). 
An example for $q=2$ and $\sigma=0.8$ is given in Figure \ref{fg:fig4}. 
The magnetization is rescaled by using the exact value for $\beta/\nu$. 
The time axis is rescaled by using the earlier calculated value of the exponent $z$ cited in Table \ref{tb:Table2}. 

The scaling fit may also be applied directly to the correlation length defined by equation (\ref{eq:defksi}), since at the criticality it should scale as
\begin{equation}
\xi_L(t) = L \;  f(t/L^z).
\label{scksi}
\end{equation}
Figure \ref{fg:fig5} gives the scaling fit for the case $q=2$ and $\sigma=0.9$. 
By $\xi_{Lmax}$ we denote the saturation value that $\xi_L$ attains according to the expression (\ref{eq:defksi}). It is proportional to the size $L$ in the
  limit of large $L$.   
% ------------ f i g u r e   -----------
% ---\begin{figure}[htb]
% ---\begin{center}
% ---\includegraphics[scale=0.45]{fig5a.eps}
% ---\centering\hspace{0.2cm}
% ---\includegraphics[scale=0.45]{fig5b.eps}
% ---\end{center}
% ---\caption{Scaled correlation length for $L=100,200,400$ ($q=2$, $\sigma=0.9$)}
% ---\label{fg:fig5}
% ---\end{figure}
% ---------------------------------------
As in the previous example, the scaling is performed with the same value of $z$ as given in Table \ref{tb:Table2}.

The agreement in both cases is very good. 
Nevertheless, these fits are generally less accurate than direct calculations from systems of large sizes.

\subsection{Case $q=3$}

For the three-state Potts model we expect to obtain different dynamical exponents. Also, we should be able to distinguish, depending of $\sigma$, two regimes, corresponding to the first- and second-order phase transition. 

The calculations were performed along the same lines and with similar parameters as for the preceding case, since the increase of the correlation length with time is very similar to that for $q=2$. 

\begin{figure}[h!!!]
\begin{center}
\includegraphics[scale=0.45]{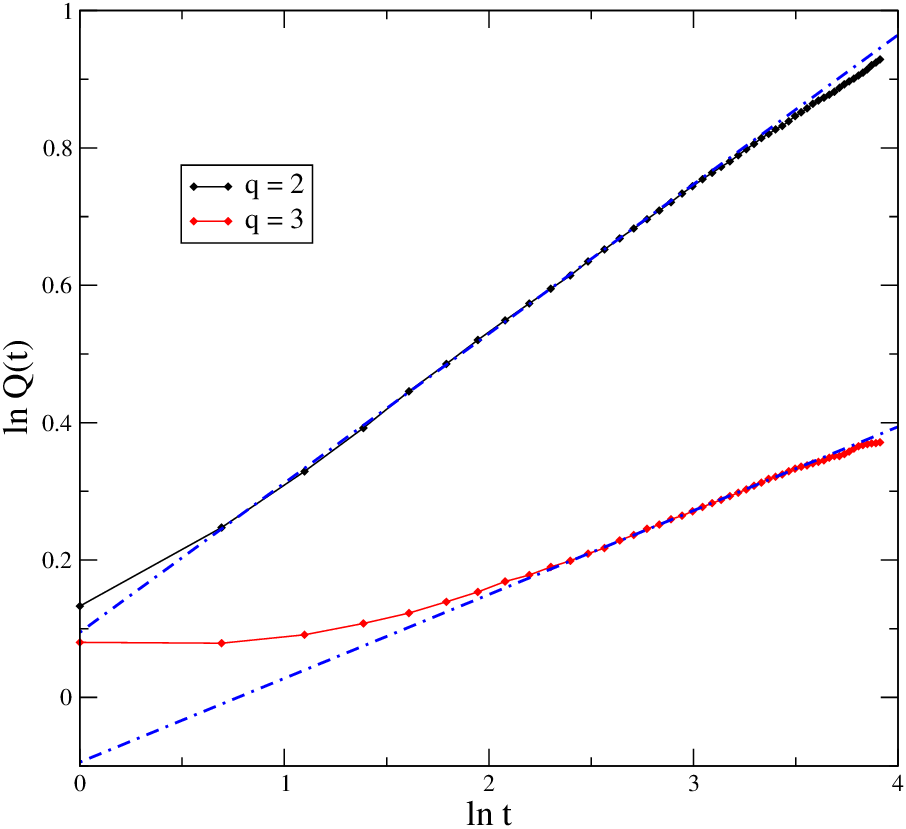}
\caption{
Comparison of $t_{micro}$ for the cases $q=2$ and $q=3$ on the example of the function $Q(t)$ (equation (\ref{eq:defQ})) for $\sigma = 0.9$, $L=3\,000$. The straight lines are the linear extrapolations performed in the scaling regime. The plot for $q=3$ was shifted by 0.5 in the y direction in order to display the two plots on the same graph.}
\label{fg:fig6}
\end{center}
\end{figure}

Yet, the microscopic time period $t_{micro}$ was found to be larger by several steps than the one for $q=2$. This property is clearly  seen in Figure \ref{fg:fig6}.
%
% ------------ f i g u r e   -----------
% ---------------------------------------
In spite of such behavior, owing to the fact that the analysis for $q=3$ is limited to larger values of $\sigma$ (as explained later), it was sufficient to use the same size $L=3\,000$ as in the $q=2$ case, but the statistics had to be increased systematically up to the 350\,000 independent runs.

\begin{figure}
\begin{center}
{\scalebox{0.45}{\includegraphics{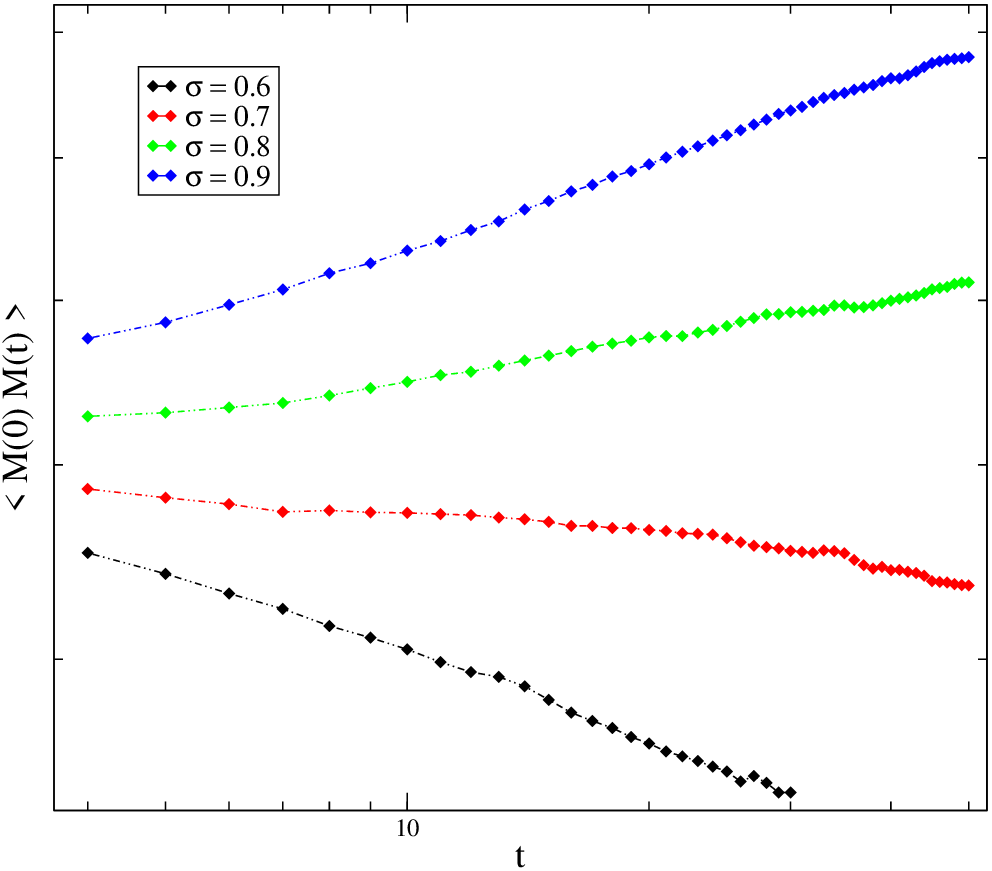}}}
\caption{The time correlation of the magnetization for $q=3$ and the selected values of $\sigma$ ($L=3\,000$).}
\label{fig7}
\end{center}
\end{figure}

In Figure \ref{fig7} we present the results for the time correlation function of the magnetization, $Q(t)$,  for the same values of $\sigma$ as in the previous case $q=2$.  
%
% ------------ f i g u r e   -----------
%\begin{figure}[h!!!]
% ---------------------------------------
The behavior is qualitatively different from the one in Figure \ref{fg:fig1}.
An initial increase is observed only for $\sigma=0.8$ and $\sigma=0.9$, while the change of behavior for lower values of $\sigma$ announces the expected onset of the first-order transition regime, which lies between $\sigma=0.7$ and $\sigma = 0.8$.
The onset of the first-order phase transition in the present model is a challenging question in itself, since the position of the tricritical point $\sigma_c(q)$ is still not known with precision \cite{BDD99,UG00,ReynalDiep04}. 
In this connection, it is important to notice, that STD has proven as an efficient approach in cases involving short-range interactions for studying both first-order phase transitions \cite{SchZ00,YZT04,YZPT06} and a the tricritical point \cite{JanssenOerding94}. 
A more detailed study of these issues in the present LR model requires a separate study \cite{GUprep}. 
We mention here only, that our preliminary results for $Q(t)$ on a finer scale of $\sigma$ locate the change of regime between 0.72 and 0.74, which is in agreement with most recent estimates \cite{ReynalDiep04} that give $\sigma_c(q=3)=0.72(1)$.
Here we shall limit the scope to the second-order transition regime analyzing further only the behavior  for $\sigma=0.8$ and $\sigma=0.9$.

The dynamical exponent $\theta'$ derived from the log-log plot of the function $Q(t)$ is given in Table \ref{tb:Table4}.
\begin{table}[htb]
\caption{The dynamical exponents for $q=3$.}\label{tb:Table4}
\begin{center} 
% \hspace*{-1.5cm}
\begin{tabular*}{8.0cm}{@{}c@{}c@{}c@{}c@{}c@{}} 
\hline\noalign{\smallskip}
$~\sigma~~$   &~~$\theta'$~~& ~~z~~&~~$z_{FSS}$~~&  ~~$\lambda/z$ \\
\noalign{\smallskip}\hline\noalign{\smallskip}
$0.9$ &~$0.120\pm .004$~~ & $1.21\pm .01 $~~ &~~$1.26\pm.04$~~ & $0.704\pm .008$ \\
$0.8$ &~$0.058\pm .004$~~ & $1.01\pm .004$~~ &~~$1.02\pm.04$~~ & $0.935\pm .006$  \\
\noalign{\smallskip}\hline
 \end{tabular*} 
\end{center} 
 \end{table}
It strongly decreases as the first-order regime approaches. 
Compared to the Ising case, the exponents $\theta'$ for $q=3$ turn out to be significantly lower.
This is similar to what was observed for the $2D$ SR Potts model, where $\theta'=0.191$ for $q=2$, while $\theta'=0.075$ for $q=3$ \cite{OSchYZ97}.

In Table \ref{tb:Table5} we also present the alternative derivation of the exponent $\theta'$ by investigating the initial slip of the magnetization for the case $\sigma=0.9$.
  \begin{table}[htb]
\caption{The exponent $\theta'$ derived from the magnetization in the case $q=3$, $\sigma=0.9$ ($L=3\,000$) for several values of the initial magnetization $m_0$ with the extrapolation to $m_0=0$.}\label{tb:Table5}
\begin{center} 
% \hspace*{-0.2cm}
  \begin{tabular*}{8.0cm}{@{}c@{}c@{}c@{}c@{}c@{}}   %     There must be rubber space between columns that can stretch to fill out the specified width
\hline\noalign{\smallskip}
$~m_0~$   &~0.1~& ~0.05~& ~0.02 ~& $m_0 \rightarrow 0$\\
\noalign{\smallskip}\hline\noalign{\smallskip}
$\theta'(m_0)$ &~$.133\pm.003$~ &~$.118\pm.004$~ &~$.112\pm .006$~ &~$.106\pm .09$\\
\noalign{\smallskip}\hline
 \end{tabular*} 
\end{center} 
 \end{table}
A rough linear extrapolation to $m_0=0$ gives a slightly smaller value for $\theta'$ than the one cited in the Table \ref{tb:Table4}. In the same time the precision of the calculations from the magnetization was considerably lower.

Within this approach, one can observe one feature common to earlier numerical calculations for the 2D short-range Potts model \cite{OSchYZ97}, that $\theta'(m_0)$ converge to the limit $m_0 \rightarrow 0$ from different sides for $q=2$ and $q=3$, which was there attributed to the opposite positions of the related fixed points.

\begin{figure}[h!!!]
\begin{center}
\centering\hspace{-1cm}
{\scalebox{0.46}{\includegraphics{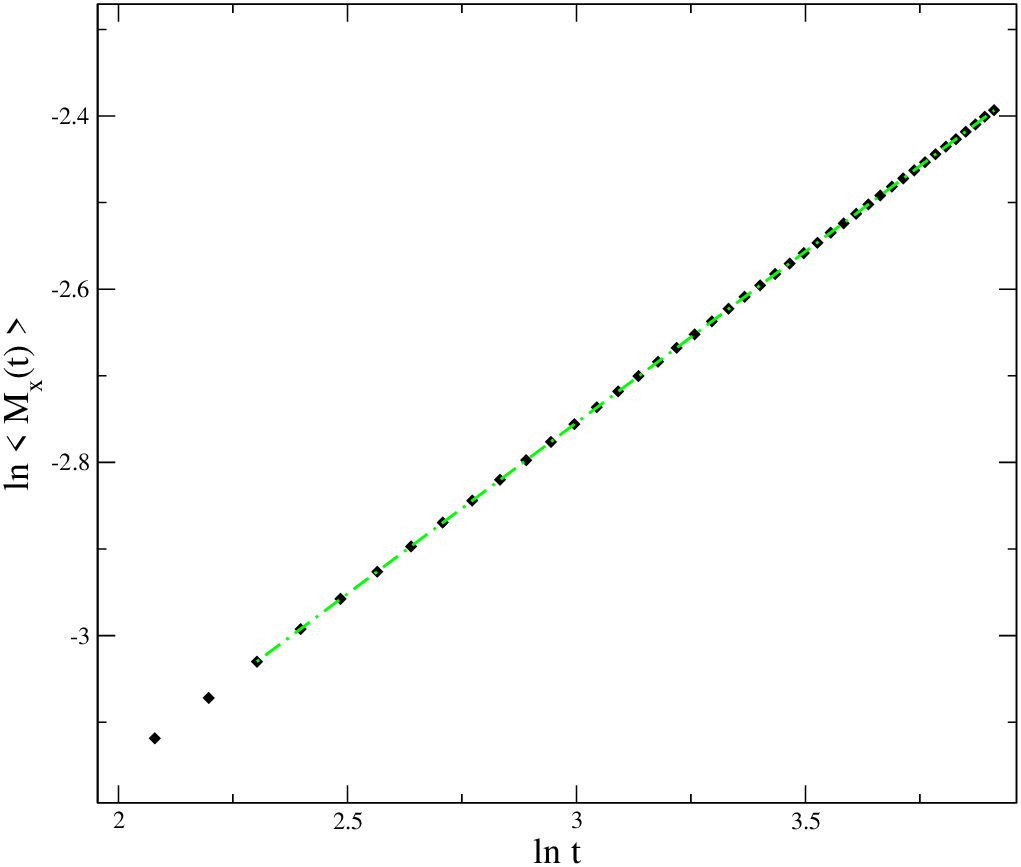}}}
\end{center}
\caption{MC data for the magnetization (\ref{eq:mxpl}) in the case $q=3$, $\sigma=0.8$, $L=3\,000$ (diamonds) and the linear extrapolation (line).}
\label{fg:fig8}
\end{figure}

The results for the exponent $z$ obtained from equation (\ref{eq:mxpl}) using the same procedure as in the case $q=2$ are presented in the third column of  Table \ref{tb:Table4} and illustrated in Figure \ref{fg:fig8} for $\sigma=0.8$. 
The values of $z$ are slightly larger than for the Ising case, similar as it was obtained for the $2D$ SR Potts model \cite{OSchYZ97}.

For want of prior results for dynamical exponents in the case $q=3$, we also applied the FSS to the  magnetization (\ref{eq:mx}) and the correlation length (\ref{eq:defksi}) by performing independent evaluations of the exponent $z$ by using the collapsing fits. The results are included in Table \ref{tb:Table4} for comparison.  

As in previous case, the results for $\lambda/z$ obtained from the autocorrelation function (\ref{autocorr}) (cited in the last column of Table \ref{tb:Table4}) agree with the independently calculated values of $\theta'$ and $z$ within the accuracy limits. 

%
% ------------ f i g u r e   -----------
% ---------------------------------------

\section{Conclusion}
 \label{sec:Conclusion}

We presented a numerical study of scaling properties related to the short-time dynamics at criticality in the 1D LR Potts model. 
Based on the analysis of several physical quantities, 
we showed that in spite of the fast relaxation in presence of the LR interactions, the STD scaling regime can be observed numerically and dynamic critical exponents evaluated with satisfactory accuracy.
We focused here on studying the problem in larger range of $\sigma$, but the accuracy of each individual result may still be improved with reasonable numerical effort. 

The dynamical exponents $\theta'$ and $z$  were evaluated in the cases $q=2$ and $q=3$, for several values of $\sigma$ belonging to the nontrivial critical regime.
The exponents are found to differ for the two cases and depend on $\sigma$, in similar way they depend on dimensionality in the SR analogue of this model.

For the Ising case, the comparison could be made with the existing RG results. 
A fair agreement for values of $\sigma$ close to the MF border ($\epsilon=2\sigma-1 \ll 1$) is obtained, but the discrepancy reaches far beyond the estimated error bars as $\sigma$ increases,
which could be attributed to the shortcomings of the $\epsilon$-expansion.
 Our results are in favor of significantly smaller increase of $\theta'$ and larger increase of $z$ with decreasing range of interactions. 

For $q=3$, new values for the exponents $\theta'$ and $z$ were obtained in, more restrained, second-order phase transition regime.
The value of the exponent $z$ is found to be slightly larger than the one for $q=2$, while increasing  number of Potts states had larger impact on the critical exponent $\theta'$ which is appreciably smaller and tends to vanish as the first-order transition regime approaches.
We also found the change in the behavior of the time correlations of the magnetization as the first-order transition sets in with lowering of the parameter $\sigma$.

Besides the onset of the first-order transition regime in this model which is already a subject of a separate study, a number of issues remain to be examined further, such as the possible effects of different dynamics, or a complementary analysis of the exponent describing the persistence probability of the global order-parameter at criticality.

%Acknowledgement: 
\begin{acknowledgement}
This work was supported by the Croatian Ministry of Science, Education and Sports through grant No. 035-0000000-3187. 
\end{acknowledgement}

% ---------------------------

\end{document}